\documentclass{aastex63}

\usepackage{graphicx,epsfig}
\usepackage{dcolumn}
\usepackage{bm}
\usepackage{xcolor}
\usepackage{amsmath}

\newcommand{\be}{\begin{equation}}
\newcommand{\ee}{\end{equation}}
\newcommand{\bse}{\begin{subequations}}
\newcommand{\ese}{\end{subequations}}
\newcommand{\bary}{\begin{eqnarray}}
\newcommand{\eary}{\end{eqnarray}}
\newcommand{\bwt}{\begin{widetext}}
\newcommand{\ewt}{\end{widetext}}

\begin{document}

\title{
Origin of sub-TeV afterglow emission from gamma-ray bursts GRB 190114C and GRB 180720B
}

\correspondingauthor{Carlos E. L\'opez Fort\'in}

\author{Sarira Sahu}
\email{sarira@nucleares.unam.mx}

\author{Carlos E. L\'opez Fort\'in}
\email{carlos.fortin@correo.nucleares.unam.mx}
\affiliation{Instituto de Ciencias Nucleares, Universidad Nacional Aut\'onoma de M\'exico, \\
Circuito Exterior, C.U., A. Postal 70-543, 04510 Mexico DF, Mexico}

\begin{abstract}
The detection of GRB 180722B and GRB 190114C in sub-TeV gamma-rays has opened up a new window to study gamma-ray bursts in high energy gamma-rays. Recently it is shown that the synchrotron and inverse Compton processes are responsible for  the production of these high energy gamma-rays during the afterglow. Here, for the first time we demonstrate that, the photohadronic scenario which is successful in explaining the multi-TeV flaring in high energy blazars is also applicable for gamma-ray bursts. We show that the sub-TeV spectra of GRB 190114C and GRB 180720B are due to the interaction of high energy protons with the background photons in the synchrotron self-Compton region and synchrotron region respectively. The nature of the background photon distributions help us to constraint their bulk Lorentz factors.
\end{abstract}


\section{Introduction} \label{sec:intro}

The standard model for the prompt emission from gamma-ray bursts (GRBs) in the $\sim$ MeV range is the {\it  fireball model}, which also predicts an afterglow in the GeV-TeV energy range when the jet runs into an external medium  and the emission can last from minutes to several hours \citep{Piran:2004ba,Kumar:2015pr}. Long after the fading away of the prompt emission, the GeV emissions were observed and had gradual temporal decay which suggests that these GeV photons were produced from the afterglow \citep{Ajello:2019avs}. Several dozen GRBs in few GeV energy range have been detected by EGRET and {\it Fermi}-LAT in the past, and multiple attempts to detect very high energy (VHE) gamma-rays ($>100$ GeV) from GRBs were unsuccessful, resulting only in upper limits \citep{Zhu:2013ufa,Totani:1998zr,Arakawa:2019cfc,Derishev:2019cgi}. However, recently, VHE photons were detected from GRB 180720B by MAGIC and GRB 190114C by HESS telescopes respectively, thus, opening a new window in the electromagnetic spectrum for the study of GRBs \citep{Acciari:2019dxz,Arakawa:2019cfc}. Observation of sub-TeV photons from the afterglow phase would also provide crucial information regarding the particle acceleration and radiation  mechanisms, leptonic and hadronic contributions to the luminosity and other microphysical parameters in GRB physics \citep{Fraija:2019rag}.

Within the afterglow scenario, it has been argued that high-energy gamma-rays above
100 MeV are produced via synchrotron mechanism in the afterglow shocks \citep{Sari:2000zp,Wang:2019zbs}.
However, it is hard to explain the sub-TeV photons detected at late times ($>100$s) as the shock is already decelerated substantially. So, beyond the synchrotron limit we must invoke other radiation mechanisms to explain the sub-TeV emission \citep{Razzaque:2009rt,Asano:2012jr,Razzaque:2010ku,Asano:2008tc}. Alternative
radiation mechanisms, such as: synchrotron self-Compton (SSC), proton synchrotron, photohadronic and the proton-proton collision processes are proposed to explain these sub-TeV emissions \citep{Band:1993eg,Uhm:2013km,Warren:2017bdm,Meszaros:1993ft,Daigne:2000xg}. The advantages and disadvantages of many models are reviewed in \citep{Kumar:2015pr}. 

The mechanisms involving hadrons suffer from poor efficiency, as these models require a much larger energy in accelerated protons than in the emitted gamma-rays \citep{Crumley:2012ra,Yacobi:2014vja}. 
On the other hand, these models are not discarded as potential sources of VHE emission, rather constraint the hadronic contribution to the jet from the non-observation of neutrino events \citep{Yacobi:2014vja}.

As discussed in the literature, a source which is active in emitting synchrotron photons, must also produce higher-energy photons through up-scattering of the  ambient synchrotron photons by the same electron population, hence multi-GeV to TeV photons are expected at early and late stages of the afterglow \citep{Derishev:2019cgi}. So, naturally in the context of leptonic model, SSC emission mechanism can be the front runner to explain the sub-TeV emissions. Recently, Wang et al. \citep{Wang:2019zbs} and Derishev et al. \citep{Derishev:2019cgi} have used this mechanism to explain the sub-TeV emission from GRB 180720B and GRB 190114C.

As the leptonic and lepto-hadronic models have a limited predictability due to large number of parameters, here we have a revised look into the hadronic model, particularly to the photohadronic one. This model is simple with minimal assumptions and is very successful in explaining the multi-TeV flaring from high energy blazars \citep{Sahu:2019lwj}. Also, as there are many similarities between blazar and GRB jets \citep{Nemmen:2012rd,Wang:2010nr}, our goal in this letter is to extend the photohadronic model to explain the sub-TeV emissions from GRB 180720B and GRB 190114C.

\section{Blazar and GRB}

Blazars, a subclass of active galactic nuclei (AGNs) and GRBs  are powered by relativistic jets from accreting black holes \citep{Urry:1995mg,Gehrels:2013xd}. While the central engines of GRBs are believed to be hyper-accreting stellar-mass black holes or rapidly spinning magnetars \citep{Woosley:1993wj}, for blazars, the central engines are supermassive black holes. 
As the jets in these objects are oriented along the observer's line of sight,
we observe them as unresolved, point-like gamma-ray sources. Due to relativistic beaming, these objects appear extremely bright and rapidly variable \citep{Abdo:2009wu}. However, the relativistic effect in GRBs is much severe than in AGNs \citep{Wu:2015opa}. 

Blazars emit electromagnetic radiation in all wavebands, from radio to gamma-rays.
Their broadband emission is non-thermal with the spectral energy distribution (SED) having two peaks \citep{Dermer:1993cz}. While the first low energy peak
is from the synchrotron emission of electrons in the jet, the second peak is generally attributed to inverse Compton scattering with the seed photons provided by the synchrotron photons around the first peak. 

The mechanism of the prompt non-thermal radiation in sub-MeV energy in GRBs is still highly debated and can be  modeled by the “Band function” \citep{Band:1993eg},
whose origin is still unknown (see, however \cite{Uhm:2013km}). Other mechanisms have also been proposed  \citep{Rees:2004gt,Peer:2005qoc,Beloborodov:2009be}. 
However, it is believed that synchrotron radiation is the leading mechanism \citep{Meszaros:1993ft,Daigne:2000xg} and is widely used.

Studies on afterglow have suggested that photons above few GeVs are difficult to interpret in terms of synchrotron mechanism, unless a large bulk Lorentz factor is employed \citep{Razzaque:2009rt} or fine tuning of the GRB parameters are considered \citep{Fraija:2019rag}. On the other hand, SSC emission had predicted the production of VHE photons in the early emission stage of afterglow \citep{Meszaros:1993ft,Beniamini:2015eaa}. So far, mostly SSC mechanisms are used to interpret the afterglow VHE emissions \citep{Kumar:2015pr,Warren:2017bdm,Derishev:2019cgi,Wang:2019zbs}.

Several studies have been undertaken to compare the blazar and GRB properties. The spectral properties of blazars and optically bright GRB afterglows were compared \citep{Wang:2010nr} and found that GRB afterglows have the same radiation mechanism as BL Lac objects. A similar correlation of the synchrotron luminosity and Doppler factor between GRBs and AGNs has been found \citep{Wu:2011xrt}. Nemmen et al. suggests that the relativistic jets in AGNs and GRBs have
a similar energy dissipation efficiency \citep{Nemmen:2012rd}. All the above studies provide evidences that the jets in GRBs and blazars are similar despite the large discrepancy in their masses and bulk Lorentz factors.

Flaring in multi-TeV seems to be a major activity of the blazars, which is unpredictable
and switches between quiescent and active states involving
different time scales and flux variabilities \citep{Senturk:2013pa}. Although, the flaring mechanism is not well understood it can be explained using leptonic and hadronic processes in the jet \citep{Boettcher:2013wxa,Cerruti:2016all}. Assuming that the photohadronic process $p\gamma\rightarrow \Delta^+$ is effective in the jet during the flaring, we have explained the multi-TeV flaring events from HBLs very well \citep{Sahu:2019lwj,Sahu:2019kfd}.

As discussed above, there is a similarity between the blazar jet and the GRB jet, it is tempting to use the photohadronic model to study the sub-TeV spectra observed from GRB 180720B and GRB 190114C. The observed VHE flux is given by
\be
F_{\gamma}(E_{\gamma}) = F_{\gamma,int}(E_{\gamma}) e^{-\tau_{\gamma\gamma}}.
\label{FluxRel}
\ee
In the observation of multi-TeV emission from the HBLs, the extragalactic background
light (EBL) plays very important role and 
the exponential factor $e^{-\tau_{\gamma\gamma}}$ in Eq. (\ref{FluxRel}) is the depletion factor from the
interaction of VHE gamma-rays with the EBL to produce the electron-positron pairs, where
$\tau_{\gamma\gamma}$ is the optical depth for the process
$\gamma\gamma\rightarrow e^+e^-$. To account for the attenuation 
of these high energy gamma-rays well known EBL models are used \citep{Franceschini:2008tp}.
The intrinsic flux in the photohadronic model is given by 
\be
F_{\gamma,int}(E_{\gamma})=F_0
\left (   \frac{E_{\gamma}}{TeV}
\right )^{-\delta+3},
\ee
where $E_{\gamma}$ is the observed energy of the sub-TeV photon. 
Here $F_0$ is the flux normalization factor which can be fixed from the observed spectrum and the spectral index $\delta$ is the only free parameter in the model.  We have observed that, there are roughly three types of flaring states depending on the value of $\delta$ and its value is constraint in the range $2.5\le \delta \le 3.0$ \citep{Sahu:2019kfd,Sahu:2019scf}. In this model $\delta=\alpha+\beta$, where $\alpha \ge 2$ is the spectral index of the Fermi accelerated proton in the jet and $\beta$ is the spectral index of the background seed photon. The kinematical condition for the $\Delta$-resonance is,
\be
E_{\gamma} \epsilon_{\gamma} = 0.032\, \Gamma\,{\cal D}\, (1+z)^{-2} \, \mathrm{GeV^2},
\label{KinemCond}
\ee
where $\epsilon_{\gamma}$ is the background photon energy, $\Gamma$ and ${\cal D}$ are the bulk Lorentz factor and the Doppler factor respectively and for GRBs $\Gamma\simeq{\cal D}$. We have shown earlier that, to fit the the observed spectrum, it is not necessary $\it a\, priori$ to know the value of $\beta$. However, we fix the value $\alpha=2$ which fixes the $\beta$ value for different emission states. In our model, the sign of $\beta$ will tell, whether the seed photons are in the synchrotron region or in the SSC region.

\section{Results}

The recent observation of two long GRBs, GRB 190114C and GRB 180720B during afterglow in sub-TeV gamma-rays has opened up the new window to study the GRBs by ground based Cherenkov telescopes. Recently, it has been shown that the observed broadband spectra from these GRBs can be explained with the synchrotron and SSC emissions of the afterglow shocks. However, for the first time, here, we explain both the VHE spectral energy distributions of the GRBs using the photohadronic scenario and the EBL correction to the observed spectra are taken into account by using the EBL model of Francehni et al. \citep{Franceschini:2008tp}.

\begin{figure}
{\centering
\resizebox*{0.8\textwidth}{0.5\textheight}
{\includegraphics{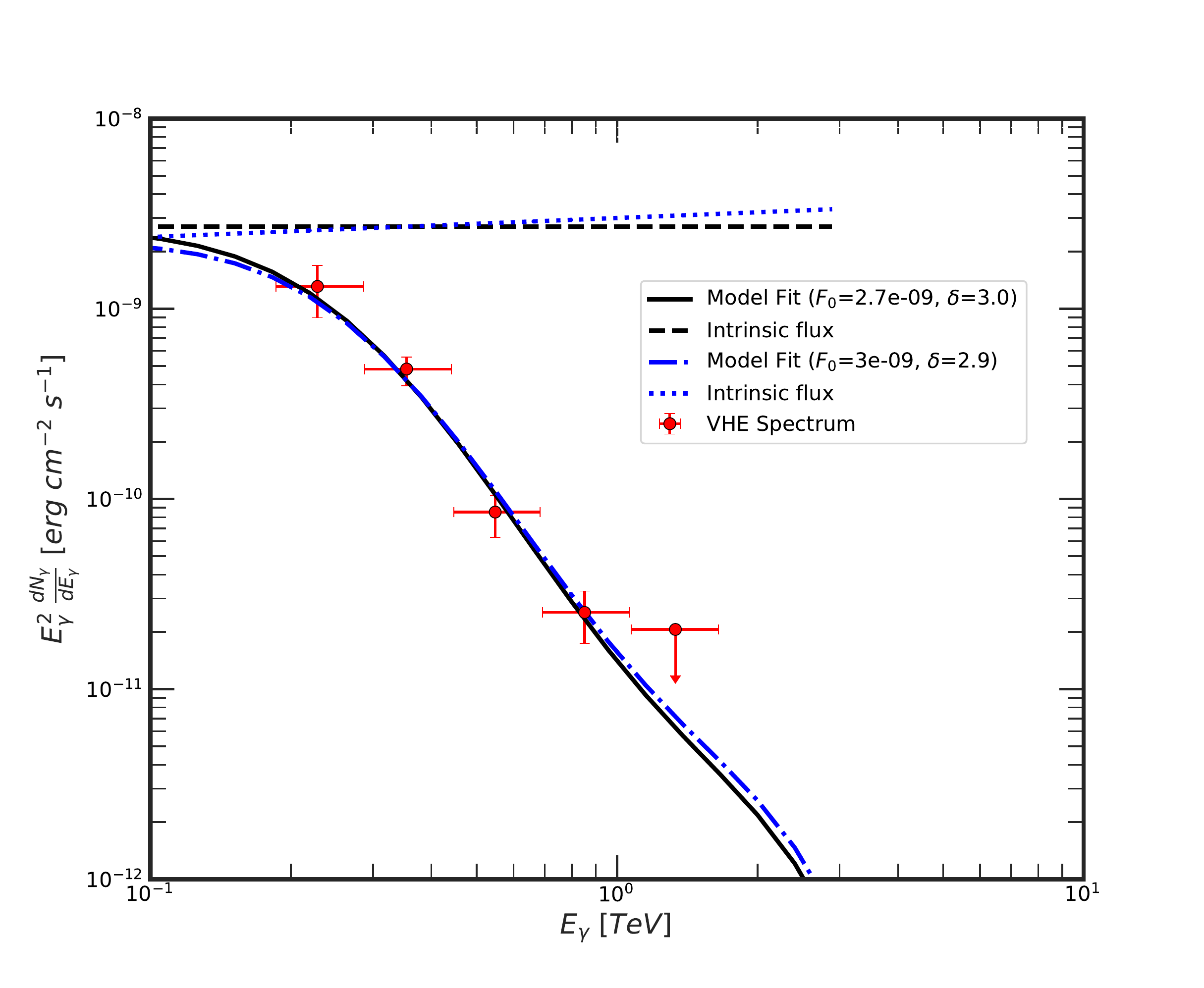}}
\par}
\caption{
The sub-TeV $\gamma$-ray spectrum observed by MAGIC telescopes from GRB 190114C \citep{Acciari:2019dxz} is fitted with photohadronic model. Two different values of $\delta=2.9$ and $3.0$ fit very will with the data and both of them are almost the same. We have also shown their corresponding intrinsic fluxes.} 
\label{fig:figure1}
\end{figure}


\begin{figure}
{\centering
\resizebox*{0.8\textwidth}{0.5\textheight}
{\includegraphics{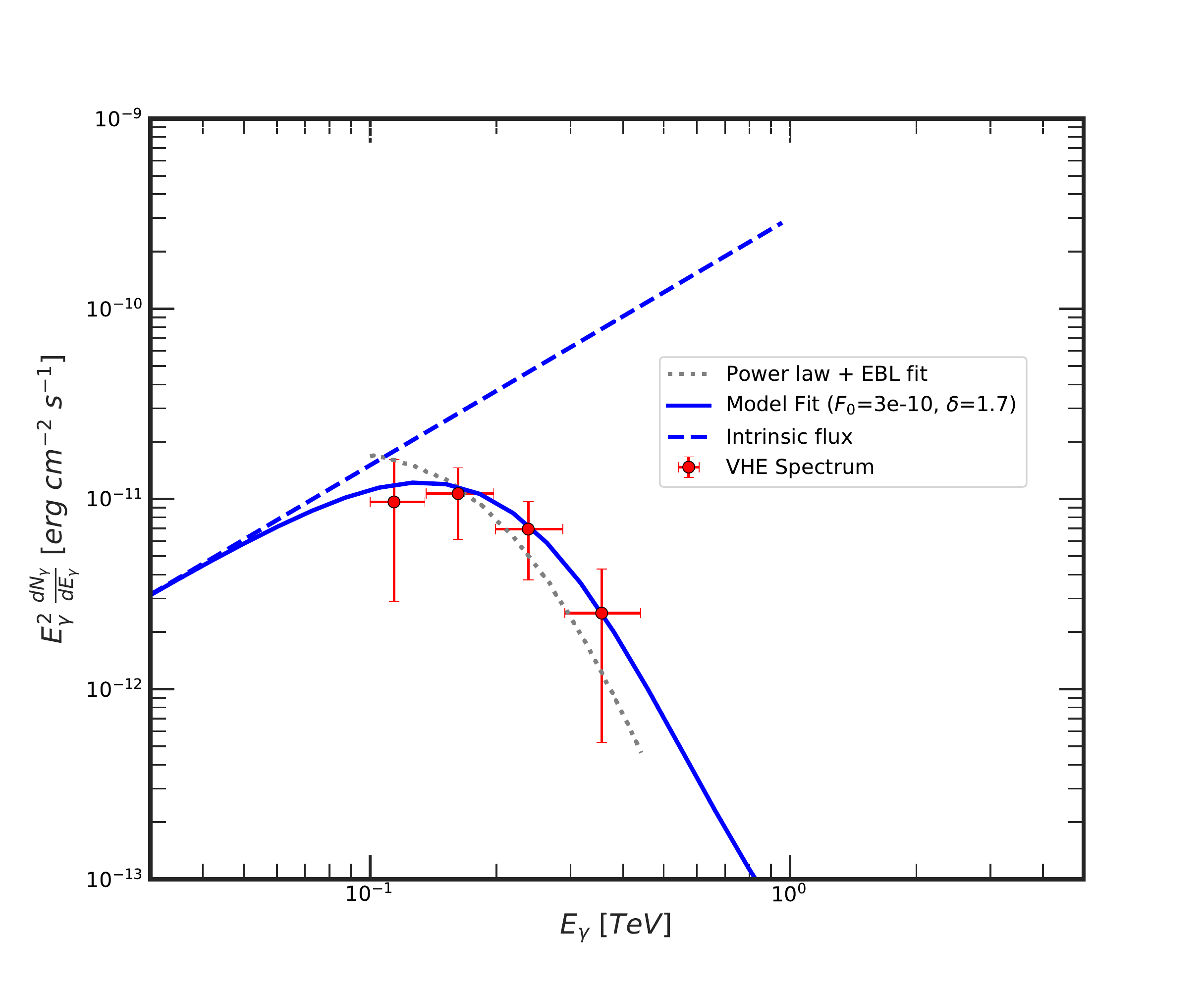}}
\par}
\caption{
The sub-TeV $\gamma$-ray spectrum observed by HESS telescope from GRB 180720B \citep{Arakawa:2019cfc} is fitted with photohadronic model. The best fit is obtained for $\delta=1.7$. The corresponding intrinsic flux is also shown. For comparison, we have also shown the power-law+EBL fit from \citep{Arakawa:2019cfc}.}
\label{fig:figure2}
\end{figure}

\subsection{GRB 190114C}

On 14th January 2019, the Burst Area Telescope (BAT) instrument on-board Swift satellite and the Gamma-Ray Burst Monitor (GBM) on board Fermi satellite  first identified the GRB 190114C as a long-duration GRB \citep{Acciari:2019dxz}. After the alert from Swift-BAT, the MAGIC telescopes slewed to the direction of  the GRB 190114C from $T_0+57$ s until $T_0+15,912$ s and detected sub-TeV photons in the energy range $0.2\, \mathrm{TeV}\le E_\gamma \le 0.9\, \mathrm{TeV}$ for the first 20 minutes with a significance of $>20\sigma$ \citep{Aleksic:2014poa,Aleksic:2014lkm}.  This is the first time a GRB was observed by MAGIC telescopes. Subsequently, it was observed by several space-based instruments in multiwavelengths and the redshift was found to be $z=0.4245\pm 0.0005$ \citep{gcn23688,gcn23708}.
Using the leptonic model, the sub-TeV emission in the early afterglow stage is explained through the SSC mechanism (\cite{Derishev:2019cgi,Wang:2019zbs}).

Using photohadronic model, we fit the VHE spectrum with (i) $\delta=3.0$ and $F_0=2.7\times 10^{-9}\, \mathrm{erg\, cm^{-2}\, s^{-1}}$ and (ii) $\delta=2.9$ and $F_0=3.0\times 10^{-9}\, \mathrm{erg\, cm^{-2}\, s^{-1}}$, which are shown in Figure \ref{fig:figure1}. Both these sets fit extremely well to the observed spectrum, but a slight difference is observed in low energy (below 100 GeV) and high energy (above 1 TeV) limits. However, according to flaring classification scheme of HBLs, set (i) corresponds to low emission state whose intrinsic flux is flat, while set (ii) corresponds to the high emission state and the intrinsic flux is proportional to $E^{-0.1}_{\gamma}$. We take the Fermi accelerated proton  spectral index $\alpha=2$, which implies the seed photon flux spectral index $\beta=1.0$ for (i) or $0.9$ for (ii). The positive value of $\beta$ corresponds to seed photon flux $\Phi(\epsilon_{\gamma})\propto \epsilon^{\beta}_{\gamma}$ or $\propto E^{-\beta}_{\gamma}$ and this is only possible for SSC photon background as in the case of HBLs \citep{Sahu:2019kfd}. As $\epsilon_{\gamma}$ is in the SSC region, it must satisfy $\epsilon_{\gamma} \gtrsim 100$ MeV, corresponding to $E_{\gamma}=852$ GeV, the highest energy $\gamma$-ray observed by MAGIC telescopes. Using this in Eq. (\ref{KinemCond}), we put a lower limit to $\Gamma \gtrsim 74$.
The high energy proton with energy $\sim 10E_{\gamma}$ will interact with the seed photons in SSC background to produce $\Delta$-resonance, which subsequently decays to $\pi^0$ and finally to $\gamma$-rays. Although, $\delta=3.0$ and $2.9$ both fit very well to the observed spectrum, here we shall consider $\delta=2.9$ as the intrinsic spectrum is a power-law proportional to $E^{-0.1}_{\gamma}$.

The integrated flux in the energy range  $0.3\, \mathrm{TeV} \le  E_{\gamma} \le 1 \,\mathrm{TeV}$ is $F_{\gamma}=2.3\times 10^{-10}\, \mathrm{erg\, cm^{-2}\, s^{-1}}$ which corresponds to the luminosity
$L_{\gamma}=1.9\times 10^{47}\, \mathrm{erg\, s^{-1}}$ and the isotropic-equivalent energy radiated during $T_0+62$ s to $T_0+2,454$ s is $E^{\mathrm{iso}}_{\mathrm{tot}}\simeq 4.6\times 10^{50}\, \mathrm{erg}$. 
The optical depth for the $\Delta$-resonance process during the afterglow is
$\tau_{p\gamma} =n'_{\gamma} \sigma_{p\gamma} R'$, where $R'\simeq 10^{18}$ cm is the comoving distance from the central engine and $n'_{\gamma}$ is the comoving background photon density. Assuming a mild efficiency of the process, we have $\tau_{p\gamma} < 1$ and this gives $n'_{\gamma} < 2\times 10^{9}\, \mathrm{cm^{-3}}$. At this moment, the $e\gamma$ interaction takes place in the same photon background, which again gives $n'_{\gamma}< 1.5\times 10^6\, \mathrm{cm^{-3}}$. Taking the upper limit of the photon density  $n'_{\gamma}< 10^7\, \mathrm{cm^{-3}}$, the $\tau_{p\gamma} \simeq 5\times 10^{-4}$, which gives the proton luminosity $L_p=2.9\times 10^{51}\, \mathrm{erg\, s^{-1}}$.

Depending on the evolutionary stages of the progenitor of a GRB, the circumburst medium can either be an uniform medium with a constant density $\rho_0$ i.e. instellar medium (ISM) or a wind driven shell where the density $\rho\propto \, r^{-2}$ \citep{Derishev:2019cgi}. The GRB jet expanding into this medium produces afterglow emission. Here we estimate the proton density of the circumburst medium and the isotropic-equivalent total energy of the jet $E^{\mathrm{iso}}_{\mathrm{tot}}$ in photohadronic model using Eqs. (1) through (3) from \cite{Derishev:2019cgi}. For our estimate, we consider typical values of the parameters, the mass loss rate from the progenitor $\dot{M}\sim 10^{-6}M_{\odot}\ \mathrm{yr^{-1}}$ and the wind velocity $v_w\sim 10^{3}\ \mathrm{km}\ \mathrm{s^{-1}}$. For GRB 190114C, we have shown that, the bulk Lorentz factor $\Gamma \gtrsim 74$. So, here we take $\Gamma=80$ and express it in terms of $E^{\mathrm{iso}}_{\mathrm{tot}}$ and after time $t$, (here the observation time $t=2454$ s is used). In the wind environment we get the distance $R\sim 1.9\times 10^{18}\ \mathrm{cm}$, the wind density $\rho\simeq 1.4\times 10^{-26}\ \mathrm{g\ cm^{-3}}$ (corresponding to a particle density $n_p\sim 8.5\times 10^{-3} \ \mathrm{cm^{-3}}$) and $E^{\mathrm{iso}}_{\mathrm{tot}}\simeq 6.8\times 10^{54}\ \mathrm{erg}$. Considering the same isotropic energy but assuming this time the ISM, the density we estimate is $ \rho_0\simeq 5.3\times 10^{-27}\  \mathrm{g\ cm^{-3}}$ ($n_p\sim 3.2\times 10^{-3} \ \mathrm{cm^{-3}}$). These estimates  (energy and medium density) are consistent with the estimations from other models \citep{Derishev:2019cgi,Wang:2019zbs,Kumar:2007av}.

\subsection{GRB 180720B}

On 20th July 2018, the GRB 180720B is one of the brightest events detected by Fermi satellites. This is also the first GRB detected by HESS at $\sim 10$ h after the trigger of the event in the energy range $100-440$ GeV \citep{Arakawa:2019cfc}. Multiwavelength follow-up observations were carried out by several telescopes and the redshift of the object was found to be $z=0.653$ \citep{gcn22996,gcn22980}. In the leptonic scenario, this multi-GeV spectrum is interpreted as the SSC emission from the afterglow shock expanding in a constant density circumburst medium.

Again using the same photohadronic model we have an excellent fit to the sub-TeV spectrum in the energy range $0.1\, \mathrm{TeV} \le  E_{\gamma} \le 0.4 \,\mathrm{TeV}$ with $\delta=1.7$ and $F_0=1.11\times 10^{-11}\, \mathrm{erg\,cm^{-3}\,s^{-1}}$. This range of $E_{\gamma}$ corresponds to Fermi accelerated proton energy in the range $1\, TeV \le E_{p} \le 4 \, TeV$. The intrinsic flux raises rapidly and is proportional to $E^{1.3}_{\gamma}$. As shown in Figure \ref{fig:figure2}, in the low energy limit (below 100 GeV), the flux decreases and the observed spectrum has also a similar trend. 
For comparison we have also shown the power-law with EBL correction fit from \citep{Arakawa:2019cfc}.

As discussed earlier, the value of $\alpha=2$
is fixed, $\delta=1.7$ corresponds to $\beta=-0.3$. In GRB 190114C, the high energy protons interact with the low energy tail region of the SSC seed photons which has $\beta$ positive. On the contrary, the multi-GeV spectrum of  GRB 180720B is fitted with negative $\beta$ value ($-\beta$) which corresponds to SED in the forward synchrotron region with a falling flux proportional to $\epsilon^{-0.3}_{\gamma}$ and these synchrotron photons are produced in the external forward shock region. So, to produce the observed multi-GeV spectrum in GRB 180720B, the Fermi accelerated protons in the energy interval $1\, TeV \le E_{p} \le 4 \, TeV$ interact with the synchrotron photons in the external forward shock region. Mostly, the protons which are accelerated to desired energies are from the forward jet. However, fraction of the swept up circum-stellar material can also be fed into the jet and can be accelerated to power-law as discussed above and interact with the seed photons to produce the multi-GeV spectrum, provided they satisfy the required conditions.

Assuming that below $100 \ \mathrm{MeV}$ $\gamma$-rays are produced by synchrotron emission we can constrain the value of $\Gamma$ from the kinematical condition which gives $\Gamma \lesssim 30$. As the sub-TeV photons were produced after $\sim 10$ h, it is obvious that the jet has slowed down considerably.

In the energy range $0.1\, \mathrm{TeV} \le  E_{\gamma} \le 0.4 \,\mathrm{TeV}$, the integrated flux is $F_{\gamma}=1.1\times 10^{-11}\, \mathrm{erg\, cm^{-2}\, s^{-1}}$ and the corresponding luminosity is $L_{\gamma}=2.1\times 10^{46}\, \mathrm{erg\, s^{-1}}$.
Assuming that the SSC process is also operative in this region with a mild efficiency, we obtain comoving photon density $n'_{\gamma}< 1.5\times 10^6\, \mathrm{cm^{-3}}$, where $R'\simeq 10^{18}$ cm is the comoving distance from the central engine after $t\sim 10$ h. Assuming $n'_{\gamma}\sim \times 10^6\, \mathrm{cm^{-3}}$, gives $\tau_{p\gamma}\sim 5\times 10^{-4}$ and the upper limit to the proton luminosity in the jet is $L_p\sim 3.1\times 10^{50}\, \mathrm{erg\, s^{-1}}$.

 To estimate $E^{\mathrm{iso}}_{\mathrm{tot}}$, the values of mass loss rate $\dot M$ and the wind velocity $v_w$ are taken to be the same as in the previous case (GRB 190114C).
 For GRB 180720B, we have shown that the bulk Lorentz factor satisfy the constraint $\Gamma \lesssim 30$. Considering a wind-like environment and taking $\Gamma\sim 20$ and $t\sim 12$ h (HESS observation began at time $T_0+10.1$ h and lasted for two hours), we get $R\sim 2.1\times 10^{18}\ \mathrm{cm}$. Similarly, we also estimate the medium density $\rho\simeq 1.2\times 10^{-26}\ \mathrm{g\ cm^{-3}}$ (corresponding to a particle density $n_p\sim 7.0\times 10^{-3} \ \mathrm{cm^{-3}}$) and $E^{\mathrm{iso}}_{\mathrm{tot}}\simeq 4.7\times 10^{53}\ \mathrm{erg}$. Considering the same isotropic energy but this time the ISM environment, we obtain $ \rho_0\simeq 4.3\times 10^{-27}\  \mathrm{g\ cm^{-3}}$ ($n_p\sim 2.6\times 10^{-3} \ \mathrm{cm^{-3}}$). Comparison of these results are consistent with other models \citep{Arakawa:2019cfc, Wang:2019zbs, Kumar:2007av}.

\section{Conclusions}

For the first time, sub-TeV gamma-rays were observed from GRB 190114C and GRB 180720B by ground based Cherenkov detectors during their afterglow emissions. It is proposed that, these sub-TeV emissions are leptonic in nature and are interpreted as the SSC emission of the afterglow shocks expanding into the ambient media. On the contrary, here, for the first time, we have shown that, the interaction of few TeV protons with the background seed photons in the synchrotron and the SSC regimes are responsible for the production of sub-TeV gamma-rays.
The emission from GRB 190114C can be interpreted as the interaction of Fermi accelerated protons with the seed photons in the SSC emission regime with its flux proportional to $\epsilon^{(0.9-1.0)}_{\gamma}$. But, the multi-GeV afterglow emission from the GRB 180720B is from the interaction of high energy protons with the seed photons in the synchrotron region whose flux is a power-law, proportional to $\epsilon^{-0.3}_{\gamma}$.
This is analogous to the multiwavelength SED of blazars. By assuming that the background photons above 100 MeV are produced from SSC process and below this energy, they have synchrotron origin, we constraint the bulk Lorentz factor for GRB  180720B to be $\Gamma\lesssim 30$ and for GRB 190114C to be $\Gamma\gtrsim 74$ respectively. 
Using these constraint on $\Gamma$, we have also estimated the isotropic equivalent total energy of the jet and the circumburst density for both the GRBs, which are consistent with other models.

Without invoking many phenomenological parameters, the photohadronic model with a single parameter $\delta$ has relatively robust predictions about the VHE spectra in the afterglow phases of GRB 190114C and GRB 180720B. It also restricts the arbitrariness of the bulk Lorentz factor. 
In future, detection of more GRB afterglows in the VHE domain at redshift $\lesssim 0.5$ by existing and forthcoming Cherenkov Telescopes will provide valuable information about the GRB physics and the VHE emission mechanism(s), a litmus test for different models.

The work of S.S. is partially supported by
DGAPA-UNAM (Mexico) Project No. IN103019. 
We are thankful to Shigehiro Nagataki for simulating discussions. 


\bibliographystyle{aasjournal}
\bibliography{grbv1.1}{}

\end{document}